\documentclass[12pt,preprint]{aastex}

\slugcomment{Submitted to Astrophysical Journal}

\shorttitle{RESIK K spectra in solar flares}
\shortauthors{Sylwester et al.}

\begin{document}

\title{Highly Ionized Potassium Lines in Solar X-ray Spectra and the Abundance of Potassium}


\author{J. Sylwester and B. Sylwester\altaffilmark{1} }
\affil{Space Research Centre, Polish Academy of Sciences, 51-622, Kopernika~11, Wroc{\l}aw, Poland}
\email{js@cbk.pan.wroc.pl}

\and

\author{K. J. H. Phillips\altaffilmark{2}}
\affil{Mullard Space Science Laboratory, University College London, Holmbury St Mary, Dorking,
Surrey RH5 6NT, U.K.}
\email{kjhp@mssl.ucl.ac.uk}

\and

\author{V. D. Kuznetsov\altaffilmark{3} }
\affil{Institute of Terrestrial Magnetism and Radiowave Propagation (IZMIRAN), Troitsk, Moscow, Russia}
\email{kvd@izmiran.ru}

\begin{abstract}
The abundance of potassium is derived from X-ray lines observed during flares by the RESIK instrument on the solar mission {\it CORONAS-F} between 3.53~\AA\ and 3.57~\AA. The lines include those emitted by He-like K and Li-like K dielectronic satellites, which have been synthesized using the {\sc chianti} atomic code and newly calculated atomic data. There is good agreement of observed and synthesized spectra, and the theoretical behavior of the spectra with varying temperature estimated from the ratio of the two {\it GOES} channels is correctly predicted. The observed fluxes of the He-like K resonance line per unit emission measure gives ${\rm log}\,A({\rm K}) = 5.86$ (on a scale ${\rm log}\,A({\rm H}) = 12$), with a total range of a factor 2.9. This is higher than photospheric abundance estimates by a factor 5.5, {\bf a slightly greater enhancement than for other elements with first ionization potential (FIP) less than $\sim 10$~eV. There is, then, the possibility that enrichment of low-FIP elements in coronal plasmas depends weakly on the value of the FIP which for K is extremely low (4.34~eV). Our work also suggests that fractionation of elements to form the FIP effect occurs in the low chromosphere rather than higher up, as in some models.}
\end{abstract}


\section{INTRODUCTION}\label{intro}

There continues to be active discussion of why and how solar coronal element abundances differ from those in the photosphere by amounts that depend on the first ionization potential (FIP). It is generally believed that elements with low FIP ($\lesssim 10$~eV) have coronal abundances that are enhanced over the photospheric abundances \citep{gre07} by factors of between $\sim 4$ \citep{fel92} and $\sim 2$ \citep{flu99}, while coronal abundances of high-FIP elements are generally within a factor 2 of photospheric abundances. Models explaining this ``FIP effect" rely on some process that separates neutral atoms from ions at the chromospheric level, since there elements with low FIP are at least partly ionized but elements with high FIP are not. Processes such as diffusion along vertical magnetic fields or magnetic fields raising ions into the corona but not neutral atoms \citep{hen97} have been considered. More recently, \cite{lam04,lam09} has proposed a model involving ponderomotive forces arising from the generation of Alfv\'en waves from the chromosphere and into coronal loops. This model has the advantage of explaining the ``inverse FIP effect" apparently present in active RS~CVn binary systems \citep{aud03}, i.e. low-FIP elements having coronal abundances that are depleted compared with photospheric abundances. The model of \cite{hen97} involving electric currents within thin flux tubes involves fractionation at the level of the temperature minimum region, and might result in elements with extremely low FIPs to have coronal abundances enhanced by factors larger than 3 or 4. \cite{fel93} has cited evidence that this is the case.

Potassium is of importance in this context as its FIP is the lowest of any common solar element: 4.34~eV. There are, however, few potassium spectral lines strong enough to be observable by ultraviolet or X-ray spectrometers. X-ray observations with a spectrometer on the {\it P78-1} spacecraft of the He-like K (\ion{K}{18}) line at 3.01~\AA\ \citep{dos85} give $N({\rm K})/N({\rm H}) = 4.2 \times 10^{-7}$ (or ${\rm log}\,A({\rm K}) = 5.62$ on a logarithmic scale where ${\rm log}\,A({\rm H})=12$), based on 6 summed spectra made up of 72 individual observations. A recent analysis of SUMER observations by \cite{bry09} uses a \ion{K}{9} line emitted by a quiet coronal region to obtain ${\rm log}\,A({\rm K}) =5.37$. Photospheric determinations give generally lower abundances: \cite{lam78} give ${\rm log}\,A({\rm K}) =5.12$, while the more recent work of \cite{tak96}, based on a non-LTE analysis of the \ion{K}{1} resonance line at 7699~\AA, gives ${\rm log}\,A({\rm K}) = 5.1$. {\bf The comprehensive tabulation of solar abundances by \cite{lod03} includes abundances from CI-type carbonaceous chondrites, for which ${\rm log}\,A({\rm K}) =5.09$, very similar to photospheric values.} Comparison of these values suggests that the potassium abundance is enhanced in coronal or flare plasmas over photospheric by a factor of between 1.8 and 3.3. This would seem to argue against a FIP mechanism such as those where the fractionation level is at the temperature minimum, resulting in much larger enhancement factors for elements such as K with very low FIPs.

The RESIK X-ray spectrometer on the {\it CORONAS-F} solar spacecraft obtained numerous spectra in the range 3.4--6.1~\AA\ in the period 2001 August to 2003 May. The instrument \citep{syl05} was a crystal spectrometer with four channels, the solar X-ray emission being diffracted by silicon and quartz crystals. Pulse-height analyzers enabled solar photons to be distinguished from those produced by fluorescence of the crystal material, enabling the instrumental background to be entirely eliminated for channels 1 and 2 (Si 111 crystal, $2d = 6.27$~\AA) and its amount accurately estimated for channels 3 and 4 (quartz $10\bar 10$ crystal, $2d = 8.51$~\AA). RESIK was uncollimated to maximize the instrument's sensitivity. Channel~1 of RESIK covers the range 3.40--3.80~\AA, and during flares the principal \ion{K}{18} lines with transitions $1s^2 - 1s2l$ ($l = s$, $p$) were observed. This paper reports on an analysis of these spectra, comparing observed and theoretical spectra. Values of the potassium abundance in flare plasmas are derived, which are compared with those expected in theories for the origin of the solar FIP effect.

\section{OBSERVATIONS}\label{obs}

Some 2795 channel~1 spectra during 20 flares were examined, ranging in {\it GOES} importance from B9.6 to X1.0, between 2002 August and 2003 February, discussed recently in a separate work \citep{phi09}. Over the period the spectra were taken, the RESIK pulse-height analyzers were close to optimum settings, enabling the solar continuum visible in channel~1 to be accurately estimated. A small level of particle radiation was eliminated using counts in ``hidden" (non-solar) bins in the spectra. The times of flares, {\it GOES} importance, and heliographic locations were given in a recent study of continuum emission \citep{phi09} and so will not be repeated here. Individual spectra were collected in data gathering intervals (DGIs) that were inversely related to the flare intensity, with typical DGIs of 2~s at the peak of strong flares and a few minutes for times when the incident X-ray flux was small. Absolute fluxes for the lines and continuum, estimated to have $\sim 20$~\% accuracy, were derived using the procedure given by \cite{syl05}. A temperature distribution should strictly be used to describe the flare conditions for each spectrum, but we have generally found that an isothermal plasma generally characterizes RESIK spectra adequately in channels~1 and 2. Although there are a number of options for the determination of the single temperature for each spectrum, we elected to use temperatures from the ratio of emission in the two channels of {\it GOES}, $T_{\rm GOES}$ using conversion factors of \cite{whi05}; the justification for this is given by \cite{phi06}.

\begin{figure}
\epsscale{.80}
\plotone{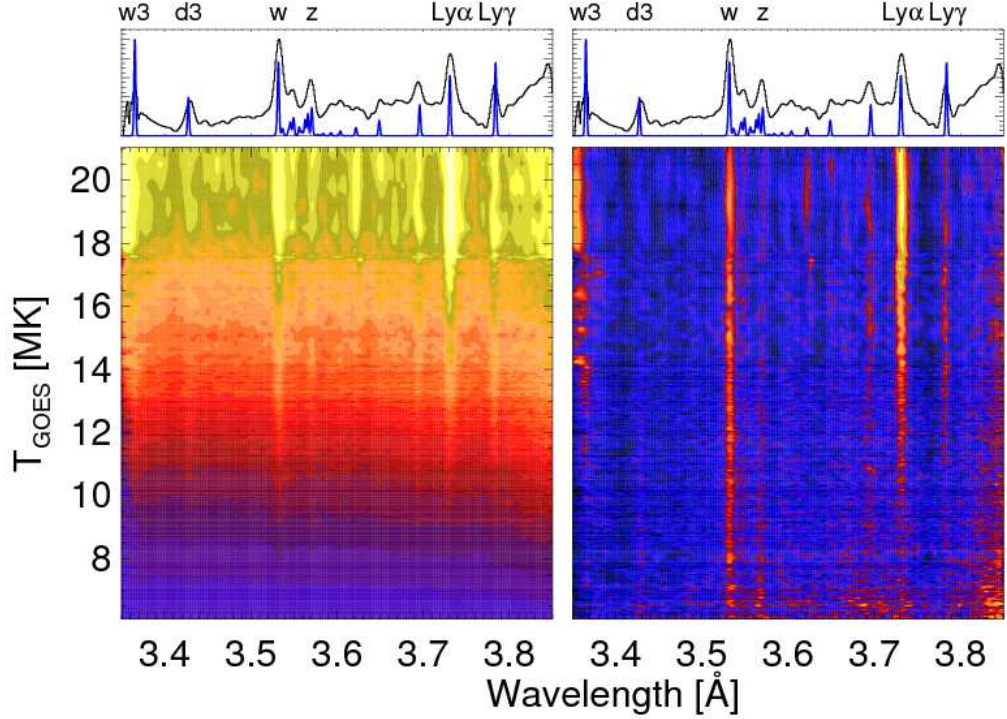}
\caption{The 2795 RESIK channel~1 spectra stacked with increasing temperature $T_{\rm GOES}$ (in MK) in the vertical direction shown in a gray-scale representation, with lighter areas indicating higher fluxes. The total spectrum is at the top of each plot, together with synthetic spectra (narrow line profiles). Principal line features are shown by letters (see Table~\ref{obs_line_list} for details). {\it Left:} Not normalized. {\it Right:} Normalized to the total number of photon counts in each spectrum. (A color version of this figure is available in the online journal.) } \label{stacked_RESIK_sp}
\end{figure}

Fig.~\ref{stacked_RESIK_sp} shows spectra stacked vertically in a gray-scale representation according to the value of $T_{\rm GOES}$. The full wavelength scale of channel~1 is shown. Without normalization (left panel), a continuous rise in both the continuum flux and flux in some lines is evident, but when the spectra are normalized by the total number of photon counts (right panel) the line emission is much clearer. The summed spectra shown at the top of the two plots indicate the principal lines. At the wavelength resolution (8~m\AA) of channel~1, three \ion{K}{18} line features are evident over the entire temperature range (4~MK--22~MK): these are the resonance ($w$) line, a blend of the intercombination lines ($x+y$), and the so-called forbidden line $z$, with emission present on the short-wavelength side of $z$ identifiable with \ion{K}{17} dielectronic satellites. Lines of \ion{Ar}{17} ($w3$, transition $1s^2 - 1s3p$, 3.666~\AA) and \ion{Ar}{18} (Ly$\alpha$, 3.731~\AA) occur at higher temperatures. This is expected from the contribution functions $G(T_e)$ (line flux per unit emission measure at electron temperature $T_e$) of these lines, which peak at $T_e = 24$~MK for the \ion{K}{18} lines but at $T_e > 30$~MK for the Ar lines. The Lyman series of \ion{S}{16} lines occurs in this region, with the Ly-$\gamma$ line at 3.784~\AA; higher members of the series are present in spectra of strong flares on the long-wavelength side of the \ion{K}{18} lines, though the series limit occurs between the \ion{K}{18} $w$ and $z$ lines. Table~\ref{obs_line_list} lists all the observable features that have been identified in RESIK channel~1 spectra with their line identifications including transitions and labels used by various authors.

\begin{deluxetable}{llll}
\tabletypesize{\scriptsize} \tablecaption{P{\sc rincipal} L{\sc ines} O{\sc bserved} {\sc in} RESIK C{\sc hannel} 1 \label{obs_line_list}} \tablewidth{0pt}

\tablehead{\colhead{Ion} & \colhead{Transition } & \colhead{Wavelength (\AA)} & \colhead{Notation}  }

\startdata

Ar XVII & $1s^2\,^1S_0 - 1s3p\,^1P_1$ & 3.366 & $w3$  \\

Ar XVI & $1s^2\,nl - 1s3p\,nl$ & 3.428 & $D3$  \\

K XVIII & $1s^2\,^1S_0 - 1s2p\,^1P_1$ & 3.532 & $w$  \\

 & $1s^2\,^1S_0 - 1s2p\,^3P_1$, $1s2p\,^3P_2$ & 3.55 & $x+y$  \\
 & $1s^2\,^1S_0 - 1s2p\,^3S_1$ & 3.571 & $z$  \\
\\
S XVI & $1s-np$ & 3.549 & Series limit \\
 \\
S XVI & $1s-7p$ to $1s-10p$ & 3.58--3.62 &Ly$\iota$--Ly$\zeta$   \\
S XVI  & $1s-6p$ & 3.649 &Ly$\epsilon$   \\
S XVI  & $1s-5p$ & 3.696 &Ly$\delta$   \\
Ar XVIII & $1s-2p$ & 3.731 &Ly$\alpha$ \\
S XVI  & $1s-4p$ & 3.784 &Ly$\gamma$ \\
\\
\enddata

\end{deluxetable}

\section{SYNTHETIC SPECTRA}\label{synth}

To compare RESIK channel~1 spectra, in particular the group of \ion{K}{18} lines at 3.53~\AA\ and 3.57~\AA, with theory, a computer program was written to calculate line fluxes for given electron temperatures $T_e$ and emission measures ($N_e^2 V$, $N_e = $ electron density and $V$ the emitting volume) and to synthesize the line and continuum spectrum over the range of channel~1. The {\sc chianti} atomic database and code (version~6) includes intensity and wavelength data for the \ion{K}{18} lines assuming direct collisional excitation from the ground state (based on distorted-wave collision strengths calculated by \cite{zha87}) as well as excitation of \ion{K}{18} lines by recombination of H-like K ions. Ionization fractions needed for this were taken from the recent work of \cite{bry09}.

{\sc chianti} does not include excitation rates for the \ion{K}{17} dielectronic satellites present near the \ion{K}{18} lines, although they make important contributions to the spectra observed by RESIK. \ion{K}{17} dielectronic satellites are formed both by dielectronic recombination of the He-like K (K$^{+17}$) ion and by inner-shell excitation of the Li-like K ion. The satellites occurring in the region of interest here have transitions like $1s^2 nl - 1s\,2p\, nl$ with $nl$ representing the ``spectator" electron. Only satellites in the $1s^2 2s - 1s2s2p$ array can be excited by inner-shell excitation for solar plasmas with densities $N_e \lesssim 10^{13}$~cm$^{-3}$. While only a small number of the dielectronically formed satellites are significant individually, the cumulative effect of spectator electrons with high $n$ is significant, particularly as their wavelengths converge on the \ion{K}{18} $w$ and $y$ lines. We therefore considered a large number of satellites with a wide variety of transition arrays. Wavelengths and intensity factors are needed to synthesize spectra. The flux at the distance of the Earth of a satellite $s$ from a solar plasma with electron temperature $T_e$ and volume emission measure $EM = N_e^2 V$ is

\begin{equation}
I(s) = 1.66 \times 10^{-16} \,\,\frac{N(K^{+17})}{N(K)} \frac{N(K)}{N(H)} \frac{F(s)\, {\rm exp}\,(-\Delta E/kT_e)}{T_e^{3/2}} \,\,\,{\rm photon}\,\, {\rm cm}^{-2} \,\,{\rm s}^{-1}
\end{equation}

\noindent where the ionization fraction $N(K^{+17})/N(K)$ is a function of $T_e$ for the flare plasma if close to ionization equilibrium, $N(K)/N(H)$ is the abundance of K relative to H, $\Delta E$ is the excitation energy of the satellite's upper level with respect to the ground level of the He-like stage, and $F(s)$ is the intensity factor for the satellite, depending on the autoionization and radiative transition rates from the satellite line's upper level (see \cite{phi94} for more details). To calculate the energy levels and the $F(s)$ factors, we ran the Hartree-Fock atomic code of \cite{cow81} with pseudo-relativistic corrections (HFR). The code in earlier forms could only be run on supercomputers, but adaptations have been made \citep{kra08} such that the code now runs on modest-sized personal computers, which was how the code was run for this case. Some 418 transitions with spectator electrons $nl$ up to $6d$ were considered. Input to the HFR code for the calculation of satellite line data includes scaling factors for Slater parameters which slightly affect the energy levels and therefore the line wavelengths. Our previous calculations \citep{phi94} have used 100\% scaling which produces line wavelengths only a few m\AA\ from observed values in the case of Ar and Fe. Unfortunately, there are no accurate observed wavelengths of \ion{K}{17} satellites available. Use of 100\% scaling for the Slater parameters in the case of \ion{K}{17} satellites produced wavelengths for high-$n$ satellites converging on 3.529~\AA\ and 3.544~\AA, about 0.0025~\AA\ shorter than the wavelengths of the \ion{K}{18} $w$ and $y$ lines. The wavelengths of all satellites were therefore adjusted in the synthesis computer program by adding 0.0025~\AA\ to the wavelengths from the Cowan program. {\bf This wavelength adjustment results in \ion{K}{17} satellite $j$ blending with \ion{K}{18} line $z$ if the line widths are due solely to thermal Doppler broadening.} This blending of $j$ and $z$ in Ca spectra was observed with the Bent Crystal Spectrometer on {\it SMM} (\cite{bel82}: note $Z({\rm Ca}) = 20$, $Z({\rm K}) = 19$), and gives further support for the way in which we adjusted the satellite wavelengths. The most important dielectronically formed satellites are listed, with their transitions, adjusted wavelengths, and $F(s)$ values, in Table~\ref{atomicdata}.\footnotemark\

\footnotetext{Full results of the calculations are available from the authors.}

The satellites in the \ion{K}{17} $1s^2\, 2s - 1s\,2s\,2p$ array are also excited by inner-shell excitation of the Li-like ion, with lines $q$ and $r$ making significant contributions to the spectrum. Excitation data for these satellites were calculated using collision strengths given by \cite{bel82} for the case of the equivalent \ion{Ca}{18} satellites, and ion fractions from \cite{bry09}.

{\sc chianti} does not give data for the higher-$n$ members of the \ion{S}{16} Lyman series, some of which are important for a  complete synthesis of the spectral region.  The generally weak \ion{S}{16} lines that are close to the \ion{K}{18} lines were included by a straight extrapolation of fluxes available from {\sc chianti} for lines with transitions $1s-np$, $n\leqslant 5$ (data from interpolation of $R$-matrix calculations by \cite{agg91}). This is justified by the fact that the collision strengths for transitions with $n=2$, 3, 4, and 5 are smoothly varying and that de-excitation of $np$ levels is almost entirely by transitions to the ground state $1s\,^2S_{1/2}$. The wavelengths of these lines were from \cite{kel87} and \cite{eri77}. Data for the \ion{Ar}{17} and \ion{Ar}{18} lines were taken from {\sc chianti}.

The continua in the range of RESIK channel~1 are principally due to free--free and free--bound emission. They were derived from {\sc chianti} routines as discussed more fully by \cite{phi09}. Free--free emission is almost entirely due to H and He, but free--bound emission is due to recombination on to the ions of a variety of elements, in particular to (in descending order of importance) Si, Fe, Mg, and O. A choice of element abundances must therefore be made, and for our purposes we chose a coronal abundance set given in the {\sc chianti} atomic database (sun\_corona\_ext.abund), based on \cite{fel92}. The ion fractions were taken from \cite{bry09}. Recombination on to fully stripped S gives rise to a small recombination edge at 3.548~\AA, between the \ion{K}{18} $x$ and $y$ lines, but our synthesis program indicates that it does not make a significant contribution to the spectrum.

{\bf Finally, in the calculation of synthetic spectra, values of the spectral line widths must be specified. For an instrument with perfect spectral resolution, these would normally be determined by thermal Doppler broadening, with temperatures in the observed range 4--22~MK of the spectra analyzed. There is generally an additional broadening due to  non-thermal motions of the emitting ions, with typical velocities of several tens of km~s$^{-1}$ for flare spectra. The combined widths amount to 2~m\AA\ (FWHM) at most. In evaluating synthetic spectra, thermal Doppler broadening only was taken into account, the line widths being dominated by the instrumental width which for RESIK channel~1 is 8~m\AA\ (FWHM).}

\begin{deluxetable}{lcccc}
\tabletypesize{\scriptsize} \tablecaption{D{\sc ielectronic} R{\sc ecombination} E{\sc xcitation of} K XVII L{\sc ines} \label{atomicdata}} \tablewidth{0pt}

\tablehead{\colhead{Transition} & \colhead{Wavelength (\AA)} & \colhead{$F(s)$ (s$^{-1}$)} &\colhead{Notation$^a$} & \colhead{Excitation energy (keV) }}

\startdata

$1s^2 4p\,\,^2P_{1/2} - 1s\,2p\,4p\,\,^2D_{3/2}$  & 3.535 & $4.42 (13)$ & & 3.246  \\
$1s^2 4p\,\,^2P_{3/2} - 1s\,2p\,4p\,\,^2D_{5/2}$  & 3.535 & $5.39 (13)$ & & 3.246  \\
$1s^2 3p\,\,^2P_{1/2} - 1s\,2p\,3p\,\,^2D_{3/2}$  & 3.539 & $1.27 (14)$ & $d15 $& 3.049  \\
$1s^2 3p\,\,^2P_{3/2} - 1s\,2p\,3p\,\,^2D_{5/2}$  & 3.540 & $1.78 (14)$ & $d13 $& 3.049  \\
$1s^2 2p\,\,^2P_{3/2} - 1s\,2p^2\,^2S_{1/2}$  & 3.553 & $6.19 (13)$ & $m$& 2.478  \\
$1s^2 2s\,\,^2S_{1/2} - 1s\,2s\,2p\,(^1P)\,^2P_{3/2}$ & 3.559 & $1.23 (13)$ & $q$ & 2.440 \\
$1s^2 2s\,\,^2S_{1/2} - 1s\,2s\,2p\,(^1P)\,^2P_{1/2}$ & 3.562 & $2.86 (13)$ & $r$ & 2.440 \\
$1s^2 2p\,\,^2P_{3/2} - 1s\,2p^2\,\,^2P_{3/2}$  & 3.563 & $4.85 (13)$ & $a$& 2.478  \\
$1s^2 2p\,\,^2P_{1/2} - 1s\,2p^2\,\,^2D_{3/2}$  & 3.566 & $2.04 (14)$ & $k$& 2.478  \\
$1s^2 2p\,\,^2P_{3/2} - 1s\,2p^2\,\,^2D_{5/2}$  & 3.569 & $2.79 (14)$ & $j$& 2.478 \\

\\
\enddata

\tablenotetext{a}{Notation of \cite{gab72, bel79}. Wavelengths are those from Cowan HFR code + 0.0025~\AA. }
\end{deluxetable}

\section{RESULTS}
\subsection{Comparison of observed and synthetic spectra} \label{Comparison}

\begin{figure}
\epsscale{.80}
\plotone{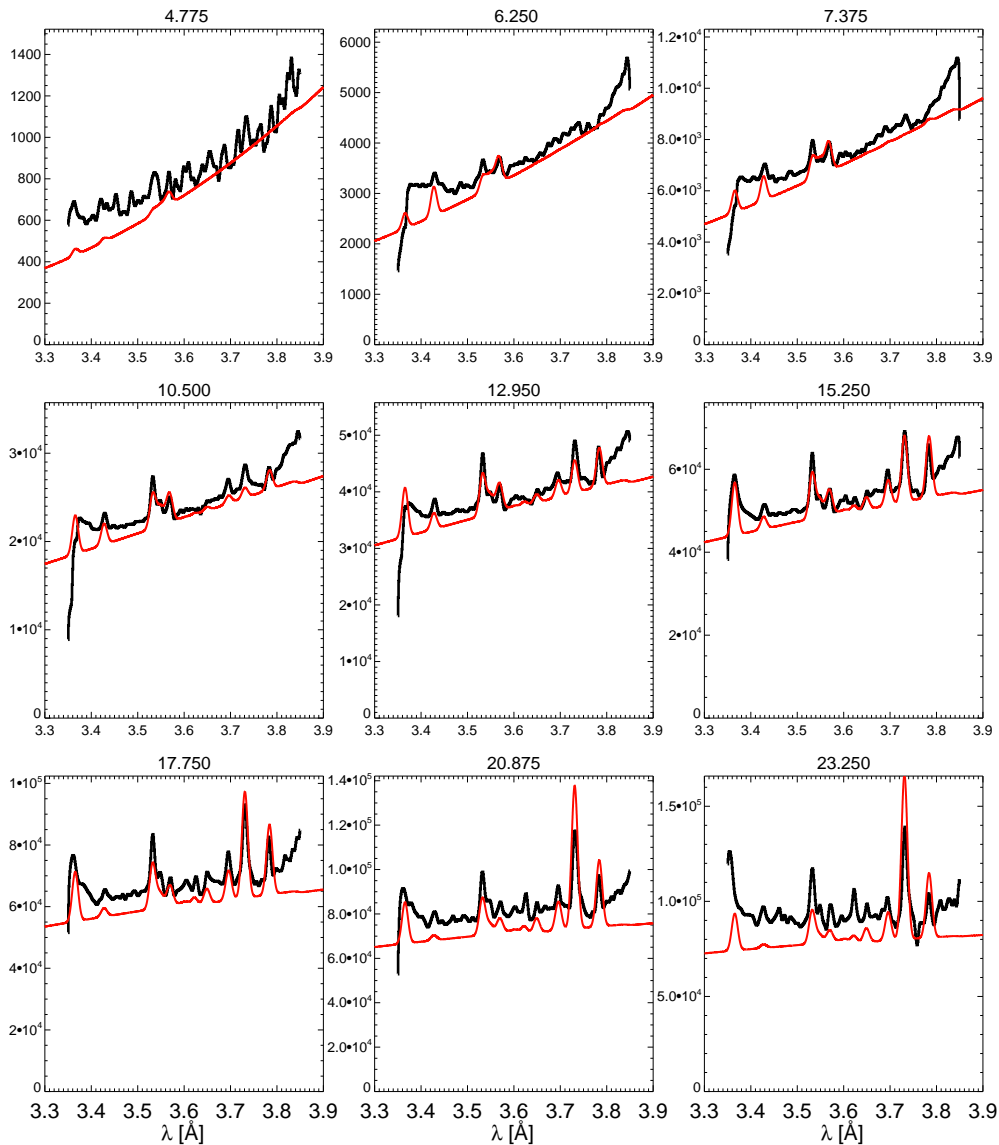}
\caption{RESIK channel 1 spectra compared with synthetic spectra for nine temperatures (values indicated above each plot in MK), with spectral irradiance (photons cm$^{-2}$ s$^{-1}$ \AA$^{-1}$) along the vertical axes. The \ion{K}{18} lines are between 3.532~\AA\ and 3.571~\AA\ (other line features are identified in Table~\ref{obs_line_list}). (A color version of this figure is available in the online journal.)}\label{comp_spectra}
\end{figure}

The observed RESIK spectra were compared with those synthesized as described in Section~\ref{synth} with temperature and emission measure from the ratio of the two {\it GOES} channels. Figure~\ref{comp_spectra} shows a selection of nine RESIK spectra from a single flare with $T_{GOES}$ over the interval 4~MK to 22~MK. For the low-temperature spectra ($T_{\rm GOES} \lesssim 10$~MK), there is a steep rise of the observed continuum emission with wavelength, very similar to the rise in theoretical free--bound and free--free emission. At these low temperatures, the slope of the theoretical continuum is extremely sensitive to temperature. Only at temperatures $\gtrsim 6$~MK is any line emission evident in the RESIK spectra. For $T_{\rm GOES} \lesssim 10$~MK, the chief lines are the \ion{K}{18} lines and the \ion{Ar}{16} $D3$ satellite feature made up of transitions $1s^2 nl - 1s3pnl$ at 3.428~\AA. {\bf The feature at the end of the range, at $\sim 3.84$~\AA, is unidentified and may not be solar in origin; it may for example be due to a curvature effect in the diffracting crystal.} At higher temperatures, the \ion{K}{18} line triplet, made up of lines $w$, $x+y$, and $z$ with blended \ion{K}{17} satellites, becomes prominent, as do the \ion{S}{16} Lyman lines and \ion{Ar}{17} and \ion{Ar}{18} lines included in Table~\ref{obs_line_list}. The observed continuum becomes larger than that in the synthetic spectra, up to 15\% by $T_{\rm GOES} = 22$~MK, which may be due to an increasing departure from an isothermal plasma. For the \ion{K}{18} lines, the $w$ line becomes steadily stronger compared with the blend of  the $z$ line with the \ion{K}{17} $j$ and $k$ satellites, as predicted by the synthetic spectra, the flux of the dielectronic satellites having an approximately $T_e^{-1}$ dependence with respect to line $w$ \citep{gab72}.

In summary, the comparison of observed and synthetic spectra shows fair agreement, especially for the \ion{K}{18} lines and nearby \ion{K}{17} satellites, which confirms our earlier results for continuum emission \citep{phi09} that a single value of temperature is valid for this narrow spectral range except for high temperatures ($T_{\rm GOES} \gtrsim 20$~MK).

\subsection{Abundance of potassium}\label{K_abundance}

\begin{figure}
\epsscale{.80}
\plotone{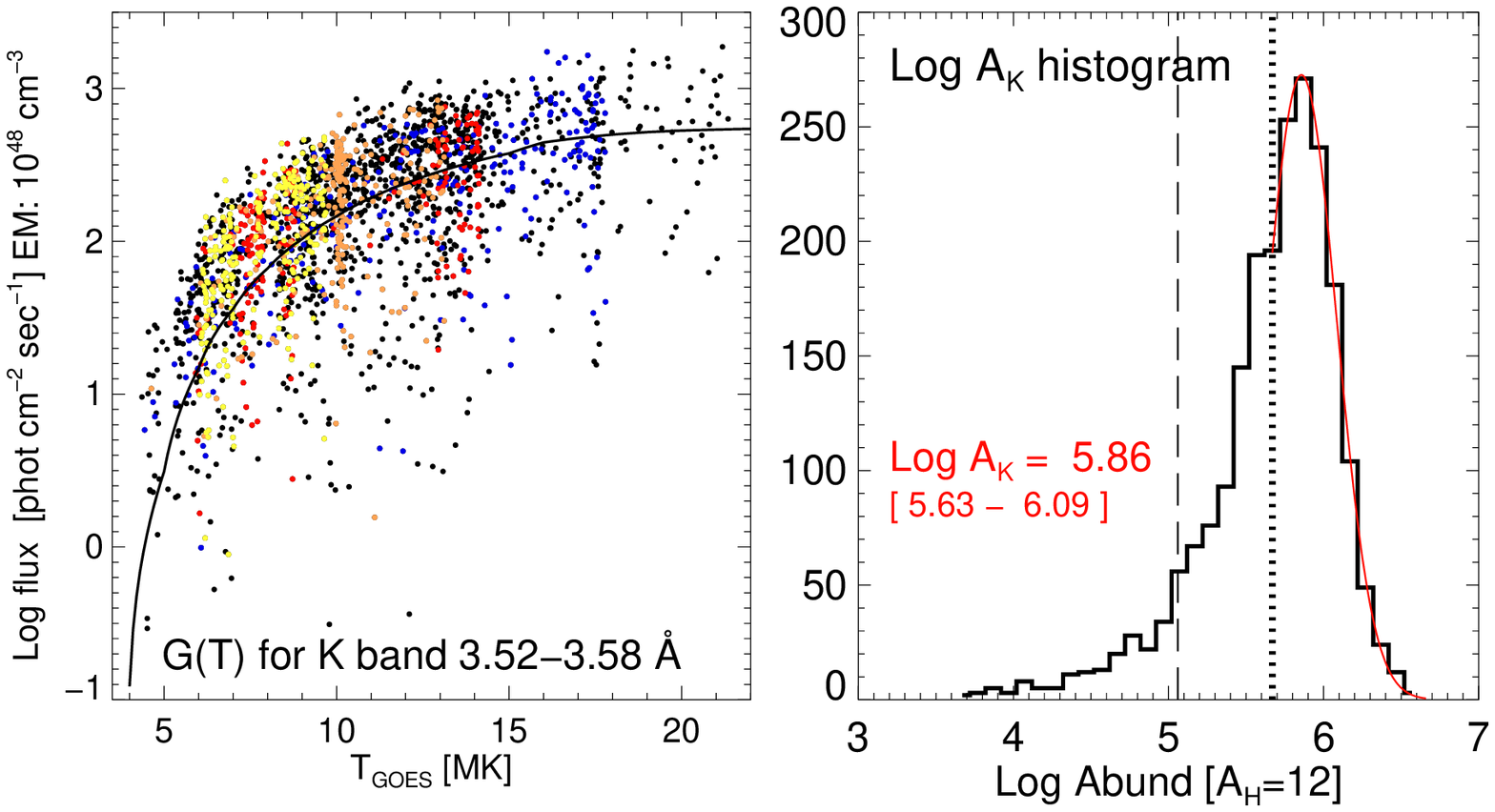}
\caption{{\it Left:} Flux of \ion{K}{18} $w$ line (including unresolved dielectronic satellites) normalized to an emission measure of $10^{48}$~cm$^{-3}$ (derived from {\it GOES}) plotted against $T_{GOES}$.  {\it Right:} Number distribution of estimates of the deduced K abundance. The peak of the distribution corresponds to a K abundance of ${\rm log}\,A({\rm K}) = 5.86$, with half-width corresponding to the range 5.63--6.09. The previous coronal (flare) abundance estimate of \cite{dos85} is shown with the dotted vertical line, the photospheric estimates of \cite{lam78} and \cite{tak96} by the dashed vertical line. (A color version of this figure is available in the online journal. In this version, points relating to individual flares are indicated by different colors to illustrate the fact that flare-to-flare variations are less than the observational scatter.) }\label{K_GofT}
\end{figure}

The RESIK observations clearly show the \ion{K}{18} lines over a wide temperature range. Using line-fitting routines, we estimated the fluxes of the line features $w$, $x+y$, and $z$ and, with temperature and emission measure derived from {\it GOES}, we derived the absolute abundance of K. The line flux for an emission measure of $10^{48}$~cm$^{-3}$ is plotted for each RESIK spectrum against $T_{GOES}$ in Figure~\ref{K_GofT} (left panel). The scatter in the 2795 points indicates the measurement uncertainties which are as expected greater for weaker, low-temperature spectra. A few outliers have small line fluxes for rather large temperatures which are mostly at the initial times of flares when the isothermal assumption with temperature given by $T_{\rm GOES}$ is least accurate. Any flare-to-flare variations, as was found from {\it Solar Maximum Mission} data \citep{syl84}, appear to be less than the scatter of the points. The theoretical curve shown is the contribution function $G(T_e)$ for the sum of the \ion{K}{18}  lines and blended \ion{K}{17} satellites for an emission measure $10^{48}$~cm$^{-3}$. The abundance of K assumed in this curve is ${\rm log}\,A({\rm K}) = 5.86$.

Figure~\ref{K_GofT} (right panel) shows the individual determinations of the absolute abundance of K from these spectra plotted as a histogram, with the logarithm of the abundance as the abscissa. These determinations were made from the departure of individual points from the theoretical curve shown in the figure. The peak of this distribution corresponds to an abundance ${\rm log}\,A({\rm K})$ of 5.86, with the half-width of the distribution corresponding to the range ${\rm log}\,A({\rm K}) = 5.63$--6.09, i.e. a factor of 2.9. This value updates the determination of \cite{phi03} (${\rm log}\,A({\rm K}) = 5.57$) based on RESIK spectra during long-duration flares with a preliminary calibration and for periods when the pulse-height analyzers were not quite at optimum operation.

This may be compared with the K abundance estimates given earlier. Our value is higher than that of \cite{dos85} by a factor 1.7, though our uncertainty range only just excludes it. It is higher by a factor of 3 than that of \cite{bry09} from SUMER measurements a quiet solar region, and is a factor 5.5 higher than the photospheric determinations of \cite{lam78} and \cite{tak96}, {\bf and meteoritic value given by \cite{lod03}}. The factor-of-2 range of our values from 20 flares covers a wide range of flares, though any flare-to-flare variations \citep{syl84} are apparently less than the scatter of the points. {\bf  Our results suggest that the coronal abundance enhancements of very low-FIP elements may increase slightly with decreasing FIP, as is suggested by the FIP model of \cite{hen97}, though the amount is too small to make a definitive comment.

This abundance determination of K and previous work relating to other low-FIP elements raises the possibility of tracing the origin of flare plasma. Assuming this plasma originates by chromospheric evaporation at the flare onset, a chromospheric origin of the plasma is expected with abundances that reflect this origin. The models proposed by \cite{lam04} and \cite{lam09} for the FIP effect were based on a fractionation level in the high chromosphere, so in the flare evaporation phase plasma with essentially photospheric origin (i.e. below the fractionation level) is delivered into the flare loops. However, there are some lines of evidence suggesting that the fractionation may occur in the lower chromosphere, below the level at which chromospheric evaporation occurs, so that coronal abundances are generally observed in flare plasmas. The present work on the abundance of K in flares appears to support the view that fractionation of elements occurs low in the chromosphere. }

\section{CONCLUSIONS}

The RESIK spectrometer has observed the He-like K ion X-ray lines in numerous flare spectra, and synthetic spectra based partly on the {\sc chianti} atomic database and partly on new calculations presented here of \ion{K}{17} dielectronic satellites compare well with the observed for temperatures and emission measures taken from the ratio of the two {\it GOES} channels. The increase of the \ion{K}{18} line $w$ relative to the nearby $z$ line at higher temperatures, due to the decreasing contribution of satellites, is evident in RESIK spectra.  Flux measurements of this line are used to derive the abundance of K for these flare plasmas with temperatures and emission measures from {\it GOES}. Comparison with theory gives a distribution of possible values of the abundance of K with peak at  ${\rm log}\,A({\rm K})= 5.86$ and half-peak range 5.63--6.09. The value 5.86 is higher than the value from X-ray flares of \cite{dos85} (by a factor 1.7) and the value from quiet-Sun SUMER spectra of \cite{bry09} (by a factor 3).  It is higher by a factor 5.5 than photospheric {\bf and meteoritic abundances of K. This indicates that the  abundance of K in flare plasmas obeys the FIP effect as discussed by \cite{fel92} and others, though there is a suggestion that the enhancement of K, the FIP of which is only 4.34~eV, is slightly more than low-FIP elements such as Mg, Si, and Fe with higher FIP. }  The factor-of-2 range of our estimates appear to be uncertainties rather than flare-to-flare variations, as has been observed for the case of Ca \citep{syl84}. {\bf In addition, our work supports the view that the fractionation of elements giving rise to the FIP effect occurs in the low chromosphere, not the high chromosphere as has been commonly proposed in the past.}

\acknowledgments

We are grateful for financial help from  the European Commission's Seventh Framework Programme (FP7/2007-2013)
under grant agreement No. 218816 (SOTERIA project, www.soteria-space.eu), the Polish Ministry of
Education and Science Grant N N203 381736,  and  the UK--Royal Society/Polish Academy of Sciences International Joint Project (grant number 2006/R3) for travel support. {\sc chianti} is a collaborative project involving Naval Research Laboratory (USA), the Universities of Florence (Italy) and Cambridge (UK), and George Mason University (USA). Dr Alexander Kramida at the National Institute of Standards and Technology is thanked for his considerable help in running the Cowan HFR code on small computers.

More information is available at
\url{http://www.aas.org/publications/aastex}.


\begin{thebibliography}{}

\bibitem[Aggarwal \& Kingston(1991)]{agg91} Aggarwal, K. M., \& Kingston, A. E. 1991, J. Phys. B, 24, 4583

\bibitem[Audard et al.(2003)]{aud03} Audard, M., G\"udel, M., Sres, A., Raassen, A. J. J., \& Mewe, R. 2003, \aap, 398, 1137

\bibitem[Bely-Dubau et al.(1979)]{bel79} Bely-Dubau, F., Gabriel, A. H., \& Volont\'e, S. 1979, \mnras, 186, 405

\bibitem[Bely-Dubau et al.(1982)]{bel82} Bely-Dubau, F., et al. 1982, \mnras, 201, 1155

\bibitem[Bryans et al.(2009)]{bry09} Bryans, P., Landi, E., \& Savin, D. W. 2009, \apj, 691, 1540

\bibitem[Cowan(1981)]{cow81} Cowan, R. D. 1981, The Theory of Atomc Structure and Spectra (Berkeley: Univ. Colorado Press)

\bibitem[Doschek et al.(1985)]{dos85} Doschek, G. A., Feldman, U., \& Seely, J. F. 1985, \mnras,
    217, 317

\bibitem[Erickson(1977)]{eri77} Erickson, G. W. 1977, J. Phys. Chem. Ref. Data, 6, 831

\bibitem[Feldman(1993)]{fel93} Feldman, U. 1993, \apj, 411, 896

\bibitem[Feldman \& Laming(2000)]{fel00} Feldman, U., \& Laming, J. M. 2000, Physica Scripta, 61, 222

\bibitem[Feldman(1992)]{fel92} Feldman, U. 1992, Physica Scripta, 46, 202

\bibitem[Fludra \& Schmelz(1999)]{flu99} Fludra, A., \& Schmelz, J. T. 1999, \aap, 348, 286

\bibitem[Gabriel(1972)]{gab72} Gabriel, A. H. 1972, \mnras, 160, 99

\bibitem[Grevesse et al.(2007)]{gre07} Grevesse, N., Asplund, M., \& Sauval, A. J. 2007, Space Sci. Rev., 130, 105

\bibitem[H\'enoux \& Somov(1997)]{hen97} H\'enoux, J. C., \& Somov, B. V. 1997, \aap, 318, 947

\bibitem[Kelly(1987)]{kel87} Kelly, R. L. 1987, {\it Atomic and Ionic Spectrum Lines below 2000 Angstroms}, J. Phys. Chem. Ref. Data, 16, Supplement 1

\bibitem[Kramida(2008)]{kra08} Kramida, A. 2008, Private communication

\bibitem[Lambert \& Luck(1978)]{lam78} Lambert, D. L.,  \& Luck, R. E. 1978, \mnras, 183, 79

\bibitem[Laming(2004)]{lam04} Laming, J. M. 2004, \apj, 614, 1063

\bibitem[Laming(2009)]{lam09} Laming, J. M. 2009, \apj, 695, 954

\bibitem[Lodders(2003)]{lod03} Lodders, K. 2003, \apj, 591, 1220


\bibitem[Phillips et al.(1994)]{phi94} Phillips, K. J. H., Keenan, F. P., Harra, L. K., McCann, S. M., Rachlew-K\"allne, E., Rice, J. E., \& Wilson, M. 1994, J. Phys. B, 27, 1939

\bibitem[Phillips et al.(2003)]{phi03} Phillips, K. J. H., Sylwester, J., Sylwester, B., \& Landi, E. 2003, \apj, 589, L113

\bibitem[Phillips et al.(2006)]{phi06} Phillips, K. J. H., Dubau, J., Sylwester, J., Sylwester, B. 2006, \apj, 638, 1154

\bibitem[Phillips et al.(2009)]{phi09} Phillips, K. J. H., Sylwester, J., Sylwester, B., \& Kuznetsov, V. D. 2009, \apj (submitted)


\bibitem[Sylwester et al.(1984)]{syl84} Sylwester, J., Lemen, J. R., \& Mewe, R. 1984, Nature, 310, 665

\bibitem[Sylwester et al.(2005)]{syl05} Sylwester, J. et al. 2005, \solphys, 226, 45


\bibitem[Takeda et al.(1996)]{tak96} Takeda, Y., Kato, K., Watanabe, Y, \& Sadakane, K. 1996, \pasj, 48, 511

\bibitem[White et al.(2005)]{whi05} White, S. M., Thomas, R. J., \& Schwartz, R. A. 2005, \solphys, 227, 231

\bibitem[Zhang \& Sampson(1987)]{zha87} Zhang, H., \& Sampson, D. H. 1987, \apjs, 63, 487


\end{thebibliography}
\end{document}